# Enhanced soliton interactions by inhomogeneous nonlocality and nonlinearity

Fangwei Ye, Yaroslav V. Kartashov, and Lluis Torner

*ICFO-Institut de Ciencies Fotoniques, and Universitat Politecnica de Catalunya, Mediterranean Technology Park, 08860 Castelldefels (Barcelona), Spain*

We address the interactions between optical solitons in the system with longitudinally varying nonlocality degree and nonlinearity strength. We consider a physical model describing light propagation in nematic liquid crystals featuring a strongly nonlocal nonlinear response. We reveal that the variation of the nonlocality and nonlinearity along the propagation direction can substantially enhance or weaken the interaction between out-of-phase solitons. This phenomenon manifests itself as a slowdown or acceleration of the soliton collision dynamics in one-dimensional geometries or of the soliton spiraling rate in bulk media. Therefore, one finds that by engineering the nonlocality and nonlinearity variation rate one can control the output soliton location.

From the viewpoint of potential application for optical signal processing and beam pointing, one of the most important properties of bright optical solitons is the particlelike behavior that they might exhibit upon interactions [1]. Depending on the character of the nonlinearity and on the interaction geometry, the interaction can have diverse outputs: For example, both one- and two-dimensional solitons can drag each other, pass through each other, preserving their identities, or fuse and give birth to new soliton signals [2]. While interactions between one-dimensional solitons and head-on collisions of their multidimensional counterparts are relatively simple, noncoplanar interaction in higher-dimensional geometries may give rise to really complex propagation trajectories. Among the illustrative examples are the spiraling of soliton pairs [3], soliton necklaces [4], and clusters [5,6] carrying nonzero net angular momentum.

Nonlinearity in suitable materials might be highly nonlocal, a property that significantly alters the propagation and interaction of light beams [7-10]. Thus, in sharp contrast to the case of local media, in nonlocal media, out-of-phase bright [11-13] and dark [14-16] solitons can attract each other, which may result in the formation of stationary [17-23] or rotating [24,25] bound soliton states. Non-coplanar soliton interactions and spiraling were recently observed in media with thermal nonlinearity with a so-called infinite range of nonlocality [26].

However, up to now soliton interactions were studied in media whose nonlocal nonlinear properties are homogeneous along the propagation direction. In this paper we address the new effects afforded by a variation of nonlocality degree and nonlinearity strength on interactions of one- and two-dimensional solitons. We discover that the interaction process can slow down (accelerate) drastically with the decreasing (increasing) nonlocality and nonlinearity. The modification of the interaction strength manifests itself in a significant variation of the output separation and of the propagation angles of interacting one- and two-dimensional solitons.

A substantial variation in the nonlocality and nonlinearity degree can be realized, e.g., in nematic liquid crystals with reorientational nonlinearity [11,27,28], when a spatially inhomogeneous (in the direction of laser beam propagation) low-frequency electric field is applied to the liquid crystal. Thermal nonlinearities afforded by lead glass, where light propagation is affected by inhomogeneous temperature distributions (arising, e.g., when the front and rear facets of the crystal are maintained at different temperatures) [26], offer an alternative setting. Here we consider a model that, under proper conditions, describes light propagation in nematic liquid crystals. We are interested in the salient physical phenomena, thus for the sake of simplicity we assume that a linearly polarized laser beam is launched at the input face of the crystal with a proper tilt that compensates the linear Poynting walk off due to the crystals birefringence [29,30]. We assume that the beam is launched far from the boundaries of the sufficiently thick liquid crystal cell. Absorption and scattering on liquid crystal inhomogeneities are neglected, on account that they are small for the length of the crystal that we consider. This assumption holds in high-quality commercial liquid crystals [11,27-30]. Under such conditions, the dynamics of laser beams in such crystals can be described by the system of coupled equations for slowly varying light field amplitude $E$ and the optically induced molecular reorientation angle $\varphi$ [31,32] as shown:

$$2ik\frac{\partial E}{\partial z} + \Delta_\perp E + k_0^2 n_a^2 \sin(2\theta_0)\varphi\, E = 0,$$

$$K\Delta_\perp \varphi - \varepsilon_a E_b^2 \frac{\sin(2\theta_0)}{2\theta_0}(1-g_0)\varphi + \frac{\varepsilon_0 n_a^2 \sin(2\theta_0)}{4}|E|^2 = 0. \qquad (1)$$

Here $k_0$ is the wave number in vacuum; $k = k_0[n_\perp^2 + n_a^2 \sin^2(\theta_0)]^{1/2}$, $n_a^2 = n_\parallel^2 - n_\perp^2$ and $\varepsilon_a = \varepsilon_0(\varepsilon_\parallel - \varepsilon_\perp)$ stand for the optical and low-frequency dielectric anisotropies, respectively, $K$ is the elastic constant, $E_b \approx V_b/D$ describes the external low-frequency electric field applied to a crystal of thickness $D$, $g_0 = 2\theta_0 \cos(2\theta_0)/\sin(2\theta_0)$, $\theta_0$ is the initial tilt of the liquid crystal molecules with respect to the beam wave vector, which can be controlled by the biasing field $E_b$, and, finally, $\Delta_\perp$ is the transverse Laplacian. When $E_b$ exceeds the threshold value $E_{th} = p/D(K/e_a)^{1/2}$ the approximation $q_0 \approx p[1-(E_{th}/E_b)^3]/2$ can be used [25]. For the sake of generality, it is convenient to transform the system (1) into the following system of equations for the dimensionless field amplitude $q$ and the nonlinear correction $n$ to refractive index:

$$i\frac{\partial q}{\partial x} = -\frac{1}{2}\Delta_\perp q - nq,$$
$$n - d\Delta_\perp n = s|q|^2. \qquad (2)$$

Here the transverse Laplacian writes $\Delta_\perp = \partial^2/\partial h^2$ or $\Delta_\perp = \partial^2/\partial h^2 + \partial^2/\partial z^2$ in the case of one- or two-dimensional geometries, respectively; the transverse $h, z$ and longitudinal $x$ coordinates are scaled to the beam width $r_0$ and the diffraction length $L_{dif} = k_0 r_0^2$, respectively; the dimensionless field amplitude is given by $q = CE$, where the parameter $C = [(n_a^4 r_0^2 k_0^2 e_0 / 4 E_b^2 e_a) q_0 \sin(2q_0)[1 - 2q_0 \cot(2q_0)]^{-1}]^{1/2}$ corresponds to the tilt angle $q_0$ taken at $x=0$, while the dimensionless nonlinear correction to refractive index is given by $n = j\, r_0^2 k_0^2 n_a^2 \sin(2q_0)/2$.

When the external electric field $E_b$ applied to the crystal becomes spatially inhomogeneous (for example, along the $x$ axis coinciding with the direction of light beam propagation), the tilt of the crystal molecules becomes a function of $x$, which, in turn, causes the nonlocality degree $d$ and the nonlinearity strength $s$ to vary *simultaneously* along the propagation direction. They are then given by

$$d(x) = \frac{K}{e_a E_b^2 r_0^2} \frac{2q_0}{\sin(2q_0)[1 - 2q_0 \cot(2q_0)]},$$
$$s(x) = \frac{n_a^4 r_0^2 k_0^2 e_0}{8 e_a C^2 E_b^2} \frac{2q_0 \sin(2q_0)}{[1 - 2q_0 \cot(2q_0)]},$$

(3)

where $q_0(x) \approx p[1 - (E_{th}/E_b)^3]/2$. It should be pointed out that, in rigor, one should take into account also a derivative $\partial^2/\partial x^2$ in the Laplacian in the second of Eqs. (2) when the biasing field is inhomogeneous along the $x$ axis. However, a careful comparison of the longitudinal and transverse scales involved shows that the coefficient in front of this term is proportional to $\sim (r_0/L_{dif})^2$; therefore, it becomes negligible for beams with widths $r_0 \gg l$ when $E_b$ varies significantly only on distances exceeding $L_{dif}$, as is the case in the settings that we address in here. The variation of $q_0$ also results in a modification of the linear refractive index $[n_\perp^2 + n_a^2 \sin^2(q_0)]^{1/2}$ along the $x$ direction. This causes only a variation in the rate of the linear phase shift acquired by the propagating beams and does not impact the dynamics of interactions between beams that is our focus in this paper.

For a typical liquid crystal with $K \approx 10^{-11}$ N, $e_a = 14.5 e_0$, $n_\| = 1.77$, $n_\perp = 1.53$, of thickness $D \approx 80$ mm, the threshold voltage amounts to $V_{th} = E_{th} D \approx 1.1$ V, while for the laser beam with width $r_0 = 4$ mm at wavelength $l = 514$ nm, the distance $x = 1$ corresponds to an actual crystal length of $\sim 0.2$ mm. A nonlocality degree $d \sim 4$ is achieved with a biasing voltage $V_b = E_b D \approx 3.1$ V. Note that solitons in such liquid crystals can be excited at mW power levels.

We assume that the voltage applied to the crystal varies linearly with $x$ and we thus introduce the dimensionless quantity $v = V_b/V_{th} = v_0 + v_x x$, where $v_0$ stands for the voltage at the input face of the crystal and $v_x \ll 1$ is the voltage variation rate. The latter is supposed to be small enough, so that the parameters of solitons adiabatically follow the modifications in the material response caused by the voltage variation. Notice that an abrupt change of the voltage applied to the crystal at a given distance can result in pronounced oscillations of the beam amplitude, which, in turn, affect interactions between several beams and may cause the modification of the output pattern in comparison with the pattern obtained for adiabatically varying parameters, even if the total variation of the voltage along the sample is the same. Figure 1 shows dependencies of nonlocality degree and nonlinearity strength on the applied voltage calculated from Eq. (3). One can clearly see that $d$ and $s$ vary with $v$ *simultaneously*, i.e., in the regime $V_b > V_{th}$ that we consider both $d$ and $s$ monotonically decrease with $v$. Moreover, the nonlinearity strength is more sensitive to the variation of the voltage than the nonlocality degree. Therefore, one may expect that the propagation dynamics of the light beam (e.g., the dependence of its amplitude and width on distance $x$) will be strongly affected by the variation in nonlinearity strength, while the nonlocality of the nonlinear response may mostly determine the character of interactions between co-propagating beams. To elucidate whether such is the case, we have conducted a comprehensive numerical analysis for different voltage gradients and input light conditions.

We first address the impact of the nonlocality and nonlinearity gradient on the interactions of one-dimensional solitons. The profiles of individual solitons were found numerically for $v \equiv v_0$ from Eq. (2) in the form $q(h,x) = w(h)\exp(ibx)$, where the function $w$ describes the soliton profile and $b$ is the propagation constant. To study soliton interactions we solved Eq. (2) with the input conditions $q|_{x=0} = w(h-h_0) + w(h+h_0)\exp(ip)$ corresponding to two identical out-of-phase solitons separated by a distance $2h_0$. In-phase soliton interactions were found to be always attractive and led eventually to soliton fusion, so we focus on out-of-phase signals. In clear contrast to the case of local media, in nonlocal media well-separated out-of-phase solitons attract each other (Fig. 2) provided that the nonlocality degree is high enough [11]. When the nonlocality and nonlinearity are homogeneous [i.e., when $v(\mathbf{x}) \equiv v_0$], such an attraction results in a periodic sequence of soliton collisions (Fig. 2(a)), caused by the fact that when the separation between solitons becomes too small attraction is replaced by repulsion. In this case the propagation distance between consecutive soliton collisions does not change with $\mathbf{x}$ and the maximal separation between soliton centers is equal to $2\mathbf{h}_0$.

This picture, however, changes drastically when the applied voltage changes with distance, which leads to the corresponding longitudinal variation of nonlocality and nonlinearity. One of the central results of this work is that varying the nonlocality and nonlinearity degree can drastically *slow down* or *accelerate* the interaction process. Figure 2(b) illustrates the dynamics of soliton interaction when the voltage increases with $\mathbf{x}$, which corresponds to decreasing nonlocality and nonlinearity. Despite the fact that nonlocality becomes weaker with distance, it still causes attraction between solitons and results in their consecutive collisions. However, the maximal soliton separation and the distance between consecutive collisions gradually increase. Physically, this effect appears as a result of a fast decrease of the nonlinearity strength that causes a broadening of each individual soliton and a decrease of the corresponding peak amplitudes (hence, a decrease of the gradients in the refractive index distribution induced by two beams). This leads to an overall reduction of the interaction forces between solitons for a fixed separation between their centers. Thus, in our setting where the nonlocality degree and the nonlinearity strength vary simultaneously with distance, but $\mathbf{s}$ changes more rapidly than $d$, the strength of nonlinearity has a central role in determining the absolute values of forces acting between solitons, the intensity and the characteristic scales at which interactions occur. At the same time, the nonlocality qualitatively changes the sign of interactions and ensures that even weak out-of-phase beams will periodically collide instead of flying apart as it occurs in local media. It should be pointed out here that the picture shown in Fig. 2(b) would be just *opposite* if only nonlocality degree would decrease with distance, while nonlinearity strength would be constant. In this case decreasing nonlocality would result in increase of peak amplitude and acceleration of interaction dynamics. Therefore, one may conclude that decreasing nonlinearity strength has an *opposite* effect on the interaction process than decreasing nonlocality, and since variation of voltage in the frames of model (1)-(3) causes much faster modification in $s$ than in $d$ (see Fig. 1), it is nonlinearity variation that determines the overall interaction dynamics. Naturally, the interaction dynamics in the case of decreasing voltage in the model (1)-(3) exhibits the opposite character: One observes a progressive reduction of the maximal separation between solitons and the distance between consecutive collisions (Fig. 2(c)), while the soliton peak amplitude increases with distance.

The above phenomenon opens the way to control the output soliton positions and to engineer the interactions by adjusting the voltage variation rate. Figure 3(a) shows the dependence of the output central positions of interacting solitons on the voltage variation rate $v_x$, at a given distance $x = 50$. This dependence is oscillatory because the number of soliton collisions that occurs at $x = 50$ depends on $v_x$: The amplitude and period of oscillations increase with $v_x$. The distance between consecutive collisions is a monotonically increasing function of the voltage variation rate (Fig. 3(b)). Our comprehensive simulations showed that to be the case at least for first five collisions. Such dependence clearly illustrates the enhancement of soliton interactions with decreasing voltage in strongly nonlocal media. A similar picture was encountered at other values of the relevant parameters, such as propagation constant, or peak amplitude, of the input solitons and the initial voltage value $v_0$.

The variation of the voltage affects in new ways the complex interactions of two-dimensional solitons, too. To illustrate this effect in typical examples, we show the outcome of non-coplanar interactions of two out-of-phase identical solitons, whose profiles were found from Eqs. (2) in the form $q(\mathbf{h},\mathbf{z},\mathbf{x}) = w(\mathbf{h},\mathbf{z})\exp(ib\mathbf{x})$. Solitons, separated by the distance $2\mathbf{h}_0$, were launched into the nonlocal medium with opposite inclination angles $\mathbf{a}_z$ and $\mathbf{p}$ phase difference, so that $q|_{\mathbf{x}=0} = w(\mathbf{h}+\mathbf{h}_0,\mathbf{z})\exp(-i\mathbf{a}_z\mathbf{z}) + w(\mathbf{h}-\mathbf{h}_0,\mathbf{z})\exp(i\mathbf{a}_z\mathbf{z})\exp(i\mathbf{p})$. The initial separation was selected in such way that a steadily spiraling soliton pair forms in the medium with fixed voltage and almost no internal pulsations arise. The existence of such rotating soliton pairs was recently reported in Ref. [24]. Here we address the evolution of such pairs caused by the variation of applied voltage. Figures 4(a) and 4(b) illustrate the dynamics encountered in the case of increasing and decreasing voltage, respectively. Mimicking the one-dimensional case, when $v_x > 0$ individual solitons forming a rotating pair gradually expand, while the peak amplitudes decrease. This is accompanied by an increase of the separation between solitons. Conversely, when $v_x < 0$ the rotating pair gradually shrinks. Importantly, besides such shape transformations, the varying voltage causes significant increase or decrease of the rotation frequency. This is in full intuitive analogy with mechanics, where conservation of angular momentum requires increase/decrease of the angular velocity with decrease/increase of the distance between the center of mass of the rotating rigid body and the pivot. To further stress this analogy, we note that the instantaneous frequency of rotation of a soliton pair can be estimated as $\Omega = L_x/M$ by analogy with the corresponding quantity in mechanics, where $L_x = \text{Im} \iint_{-\infty}^{\infty} q^* \left( h \frac{\partial q}{\partial z} - z \frac{\partial q}{\partial h} \right) dh dz$ is the $x$-projection of the angular momentum, and we introduced the "momentum of inertia" for soliton pair as $M = \iint_{-\infty}^{\infty} |q|^2 (h^2 + z^2) dh dz$. While the angular momentum $L_x$ is a conserved quantity of Eqs. (2), the momentum of inertia increases with growing separation between the soliton centers, which causes the corresponding decrease of the instantaneous rotation frequency. Interestingly, the above estimate of the instantaneous rotation frequency is rather accurate. For example, for the set of parameters corresponding to Fig. 5(a), the estimated total rotation angle of the soliton pair at $x = 50$, defined as $f = \int_0^{50} \Omega dx$, amounts to approximately 4.4, which is close to the value 3.6 obtained by direct numerical integration.

Figure 5(a) shows the angle of rotation of soliton pairs versus distance, calculated numerically for three different $v_x$ values. Notice that the slope of each curve determines the frequency of rotation at each propagation distance. While the rotation frequency is constant when $v_x = 0$, it was found to decrease slowly at $v_x > 0$, and increase when $v_x < 0$. The angular acceleration was also found to increase (decrease) with propagation distance for $v_x < 0$ ($v_x > 0$). These phenomena were found to be robust, i.e. white spatial noise added to the input distribution does not substantially affect the rotation dynamics. Figures 5(b) and 5(c) show the dependence of the rotation angle $f$ and radius $R$ on the voltage variation rate at the distance $x = 50$. The plots show that the rotation angle monotonically decreases, while the radius increases with $v_x$. The possibility to control the output soliton positions, the propagation angles, and the rotation velocity of the soliton by engineering the voltage variation rate are thus readily apparent.

Summarizing, we addressed head-on collisions of one-dimensional solitons and more complicated noncoplanar spiraling interactions of two-dimensional solitons in the media with varying nonlocality and nonlinearity degree. We focused on a physical model of a liquid crystal subject to a longitudinally varying voltage, in a model that captures the central physical phenomena. Our central finding is that adjusting the nonlocality/nonlinearity variation rate allows enhanced or weakened interaction forces between solitons drastically, affording a new way to soliton control. Our predictions can be checked experimentally with readily available liquid crystals and input light conditions.

This work has been partially supported by the Government of Spain through grant TEC2005-07815 and by the Ramon-y-Cajal program.

# Figure captions

Figure 1.   Nonlocality degree (a) and nonlinearity strength (b) versus applied voltage.

Figure 2 (color online).   Interaction dynamics of the out-of-phase one-dimensional solitons in nonlocal nonlinear medium with $v_x = 0$ (a), 0.0073 (b), and $-0.0073$ (c). Input solitons correspond to $b=2$ and $v_0 = 2.832$, while initial separation is $2h_0 = 4$.

Figure 3 (color online).   (a) Positions of output solitons at $x = 50$ versus voltage variation rate. (b) Distance $x_1$ between first and second collisions and distance $x_2$ between second and third collisions versus voltage variation rate. In all cases input solitons correspond to $b=2$, $v_0 = 2.832$, and initial separation $2h_0 = 4$.

Figure 4 (color online).   Spiraling of two out-of-phase two-dimensional solitons in nonlocal nonlinear medium with $v_x = 0.0073$ (a) and $v_x = -0.0073$ (b). Input solitons correspond to $b=1$ and $v_0 = 2.832$. The input phase tilt $a_z = 0.1$ and separation between solitons $2h_0 = 4$. Arrows indicate rotation direction.

Figure 5 (color online). (a) Rotation angle versus propagation distance for $v_x = -0.0073$ (curve 1), 0 (curve 2), and 0.0073 (curve 3). Rotation angle (b) and radius (c) at $x = 50$ versus voltage variation rate. In all cases input beams correspond to $b=1$, $v_0 = 2.832$, while input separation $2h_0 = 4$ and phase tilt $a_z = 0.1$.

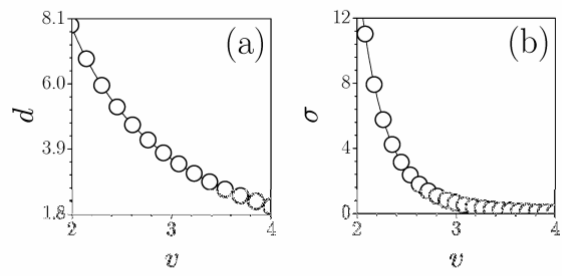

Figure 1. Nonlocality degree (a) and nonlinearity strength (b) versus applied voltage.

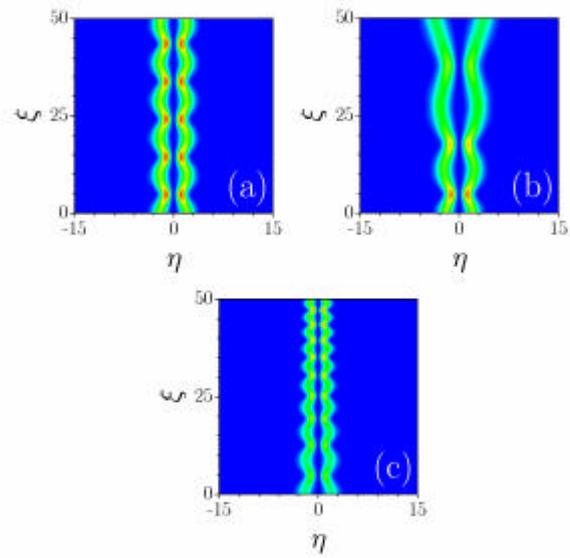

Figure 2 (color online). Interaction dynamics of the out-of-phase one-dimensional solitons in nonlocal nonlinear medium with $v_\xi = 0$ (a), $0.0073$ (b), and $-0.0073$ (c). Input solitons correspond to $b = 2$ and $v_0 = 2.832$, while initial separation is $2\eta_0 = 4$.

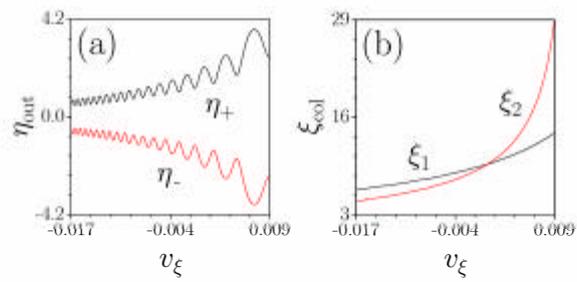

Figure 3 (color online). (a) Positions of output solitons at $\xi = 50$ versus voltage variation rate. (b) Distance $\xi_1$ between first and second collisions and distance $\xi_2$ between second and third collisions versus voltage variation rate. In all cases input solitons correspond to $b = 2$, $v_0 = 2.832$, and initial separation $2\eta_0 = 4$.

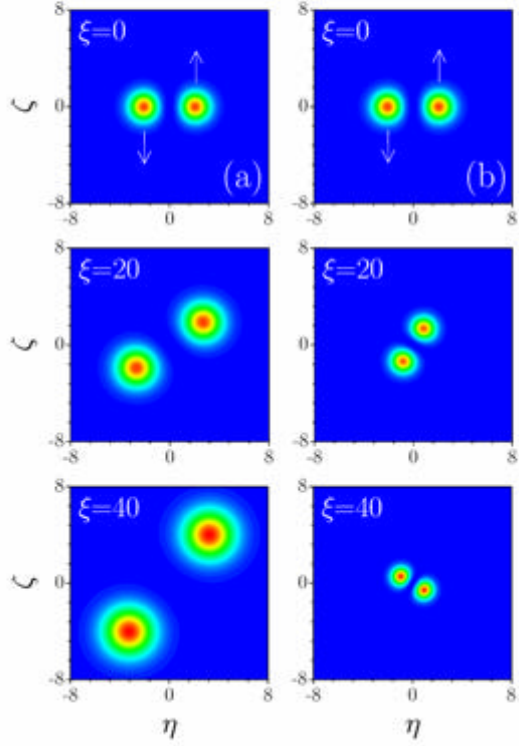

Figure 4 (color online). Spiraling of two out-of-phase two-dimensional solitons in nonlocal nonlinear medium with $v_\zeta = 0.0073$ (a) and $v_\zeta = -0.0073$ (b). Input solitons correspond to $b = 1$ and $v_0 = 2.832$. The input phase tilt $\alpha_\zeta = 0.1$ and separation between solitons $2\eta_0 = 4$. Arrows indicate rotation direction.

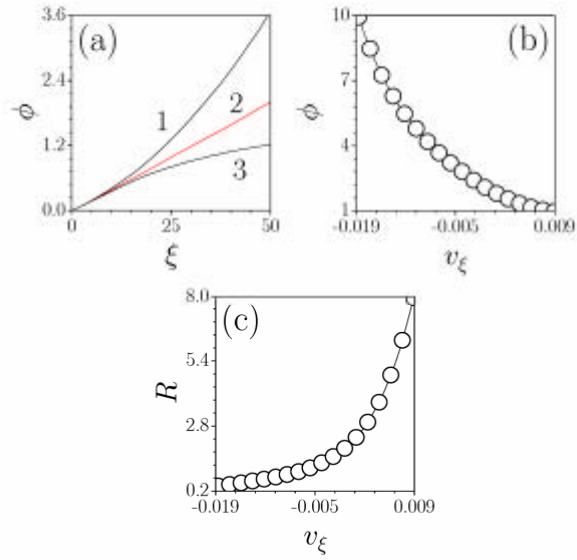

Figure 5 (color online).  (a) Rotation angle versus propagation distance for $v_\xi = -0.0073$ (curve 1), $0$ (curve 2), and $0.0073$ (curve 3). Rotation angle (b) and radius (c) at $\xi = 50$ versus voltage variation rate. In all cases input beams correspond to $b = 1$, $v_0 = 2.832$, while input separation $2\eta_0 = 4$ and phase tilt $\alpha_\zeta = 0.1$.